# To Relay or Not to Relay: Learning Device-to-Device Relaying Strategies in Cellular Networks

Nicholas Mastronarde, Viral Patel, Jie Xu, Lingjia Liu, and Mihaela van der Schaar

*Abstract-* We consider a cellular network where mobile transceiver devices that are owned by self-interested users are incentivized to cooperate with each other using tokens, which they exchange electronically to "buy" and "sell" downlink relay services, thereby increasing the network's capacity compared to a network that only supports base station-to-device (B2D) communications. We investigate how an individual device in the network can learn its optimal cooperation policy *online*, which it uses to decide whether or not to provide downlink relay services for other devices in exchange for tokens. We propose a supervised learning algorithm that devices can deploy to learn their optimal cooperation strategies online given their experienced network environment. We then systematically evaluate the learning algorithm in various deployment scenarios. Our simulation results suggest that devices have the greatest incentive to cooperate when the network contains (i) many devices with high energy budgets for relaying, (ii) many highly mobile users (e.g., users in motor vehicles), and (iii) neither too few nor too many tokens. Additionally, within the token system, self-interested devices can effectively learn to cooperate online, and achieve over 20% higher throughput on average than with B2D communications alone, all while selfishly maximizing their own utilities.

## I. INTRODUCTION

D2D communications has recently emerged as a candidate technology for supporting cooperative relaying in mobile broadband cellular networks. By taking advantage of user diversity in space and time, cooperative relaying can simultaneously increase network capacity and decrease delay [1], which can improve the performance of important applications such as wireless video streaming [3][4]. In addition to the aforementioned benefits, D2D relaying is appealing because it does not increase infrastructure costs for the network operator. This is in contrast to infrastructure relay nodes, which are included as a part of the current LTE-Advanced standard [5].

Although there are many benefits associated with D2D relaying, it requires coordination across multiple

The work of Mihaela van der Schaar and Jie Xu was supported by NSF grant no. 1218136.
N. Mastronarde is with the Department of Electrical Engineering, University at Buffalo, Buffalo, NY 14260 USA (e-mail: nmastron@buffalo.edu).
M. van der Schaar and J. Xu are with the Department of Electrical Engineering, University of California, Los Angeles, Los Angeles, CA, 90095-1594 USA (e-mail: mihaela@ee.ucla.edu and jiexu@ucla.edu).
L. Liu is with the Department of Electrical Engineering and Computer Science, University of Kansas, Lawrence, KS 66045 USA (lingjialiu@ittc.ku.edu).

nodes in the network, which is especially difficult to realize when nodes are cell-phones, laptops, or tablets (hereafter, *User Equipments*, or UEs) that are owned by *self-interested* users, who aim to maximize their own utilities. In particular, most existing work on cooperative communications/relaying implicitly assumes that all network devices are *obedient*, i.e., they will act as relays whenever requested. In practice, however, since relaying costs energy and provides no tangible benefit to the relay, obliging UEs to serve as relays could quickly lead to widespread user dissatisfaction and, ultimately, loss of customers and revenue for the network operator (for instance, a UE's battery can quickly drain while relaying data for other users, making it so that the owner cannot use her UE when she needs it). Indeed, according to the J.D. Power and Associates 2012 U.S. Wireless Smartphone Customer Satisfaction Study [6], poor battery life reduces both customer satisfaction and loyalty. Therefore, to realize the potential of D2D relaying in cellular networks, a system must be developed that incentivizes UEs to cooperate while allowing them to act as relays only when it is in their self-interest.

*A. Related Work*

There is a lot of great literature on cooperative relaying in wireless and cellular networks, e.g., [2]-[4][7][8], including work on D2D relaying [9][10][11]. However, this work does not consider self-interested users who require incentive to cooperate. Nevertheless, many of the techniques and solutions in this literature can be implemented in conjunction with different incentive mechanisms; in this sense, work on cooperative/D2D relaying techniques is complementary to work on incentivizing relaying.

Various mechanisms have been proposed to incentivize relaying in ad-hoc, cellular, and peer-to-peer networks. These mechanisms can be roughly grouped into three classes: bandwidth exchange mechanisms, reputation-based mechanisms, and token-based mechanisms.

In **bandwidth exchange mechanisms**, when device *A* receives relay service from device *B*, device *A* immediately delegates a portion of its bandwidth to device *B* as compensation for relaying [12][13]. Variants of bandwidth exchange mechanisms have been proposed based on exchanging transmission time [14] or directly exchanging relay services [15]. Unfortunately, these mechanisms are suboptimal because they require device *A* to *immediately* and *directly* reciprocate to device *B*, even though device *B* may not need assistance at the current time or, importantly, may not need assistance from device *A*. Meanwhile, token-based mechanisms and many reputation-based mechanisms exploit *indirect reciprocity* [16], which is based on the idea that device *C* will relay for device *B* because device *B* relayed for device *A* in the past. Due to this flexibility, indirect reciprocity can achieve better network performance than direct reciprocity.

**Reputation-based mechanisms** enable devices to identify noncooperative devices so that they can be eliminated as potential relays and/or punished by dropping their packets [16][19][20][21]. To identify noncooperative devices, each device tracks the other devices' reputations based on its first-hand interactions [21] and, in some implementations, also uses second-hand experience obtained through reputation propagation [16][19][20]. Unfortunately, reputation-based mechanisms are not suitable in cellular networks for two reasons. First, they do not scale well to large-scale cellular networks because first-hand experience must be tracked for many devices and reliable second-hand reputation information cannot be propagated in a timely fashion. Second, most reputation-based mechanisms rely on the broadcast nature of wireless networks to enable devices to monitor each other's transmissions [16][19][20]; however, omnidirectional broadcasting will become less common in major 4G mobile broadband cellular standards [17][18] because they support MIMO and beamforming.

*Token-based mechanisms* have been proposed to incentivize cooperation in ad-hoc [22][23][24][25][26], cellular [27][28], and peer-to-peer networks [29]. Electronic tokens work as a virtual currency: users pay tokens to other users in exchange for providing services. Although tokens have been used to incentivize cooperation in cellular networks [27][28], optimal and practical solutions are still far from being developed. For instance, [27] focuses on developing a suitable system model and architecture, but assumes that each UE deploys an arbitrary fixed threshold cooperation policy, which does not depend on its experienced network environment, thereby resulting in suboptimal performance.

In our prior work [28], we use a token system to incentivize UEs to provide relay services to each other. In particular, we focus on how a system planner can design a token system *offline* to maximize its *efficiency* (i.e., the probability that UEs that need help from a relay will get it) within an idealized network operating in equilibrium. We show that, if the expected future benefit of having an additional token outweighs the immediate cost of relaying, then a self-interested UE will be willing to act as a relay; we prove that threshold strategies are the only strategies that a self-interested UE, which wants to maximize its own utility, will adopt; and we determine the optimal token supply that should be deployed in the network to maximize the system efficiency. This theory is further developed in [30]. However, in [28][30], we do not address how individual UEs can dynamically optimize their cooperation policies in a non-idealized network environment, operating out of equilibrium, which is of paramount importance in real systems.

*B. Our Contributions*

In this report, we adopt a token-based approach similar to [28]. However, rather than focusing on incentive design from the planner's perspective, we investigate how the UEs can *learn* to adapt their cooperation strategies online, which they use to decide whether or not to provide downlink relay services for other devices in exchange for tokens. Specifically, we study how an *individual UE* in the network can *learn* its optimal[1] cooperation policy online based on its experienced network environment, and how this strategy impacts and is impacted by the other UEs in the network, which are simultaneously learning. Our contributions are as follows:

---

[1] We mean *optimal* in the *best response* sense [38]: that is, a UE's optimal cooperation policy maximizes its long-run utility given its experienced network environment and the aggregate behavior of the other UEs.

- We formulate the decision problem faced by each UE, namely, the problem of deciding whether or not to relay, as a Markov decision process (MDP). Each UE's objective is to maximize its long-term utility, which is defined as the difference between (i) the benefits it gains over time by receiving data through a relay and (ii) the energy costs it incurs over time by relaying data for other UEs.

- We formulate the MDP such that a UE's optimal cooperation policy not only depends on its current geographic location, its distance from the nearest base station, and the channel conditions, but also on the other UEs' locations and cooperation strategies. In other words, each individual UE's optimal cooperation policy is coupled with both the network environment and the other UEs.

- We assume that each UE has a relay energy budget that specifies how much energy it is willing to consume relaying data for other users. We consider a UE's remaining relay energy budget when determining its optimal cooperation policy.

- We show experimentally that optimal cooperation strategies are threshold in a UE's token holding state, and that the threshold decreases as the UE's relay energy budget decreases.

- We propose a simple, low-complexity, supervised learning algorithm that each UE can deploy to learn its optimal cooperation policy online. In the proposed algorithm, a UE estimates two parameters from its interactions with the network environment, i.e., the frequency with which it is asked to relay data for other UEs and the frequency with which other UEs relay data for it. Using its estimates of these two parameters, the UE selects its current cooperation policy from a pre-computed look-up table.

- We systematically evaluate the proposed learning algorithm in various deployment scenarios involving users with high and low mobility and UEs with high and low relay energy budgets. Our simulation results suggest that UEs have the greatest incentive to cooperate when the network contains many UEs with high relay energy budgets, many highly mobile users (e.g., users in motor vehicles), and neither too few nor too many tokens.

The remainder of this report is organized as follows. In Section II, we present the system model. In Section III, we formulate an individual UE's optimization problem. In Section IV, we propose the supervised learning algorithm that each UE can deploy to learn its optimal cooperation policy online. In Section V, we present our simulation results. Finally, we conclude the report in Section VI.

## II. SYSTEM MODEL

*A. Downlink Network Model*

In this section, we present our downlink network model. For illustration, we use an OFDM-based network model similar to, e.g., LTE/LTE-Advanced or IEEE 802.16 (WiMax); however, the optimization and learning solutions proposed in Sections III and IV, respectively, can be applied in other networks such as 3G or ad-hoc networks. We consider a cellular network containing $N$ mobile transceiver devices, which we refer to as User Equipments (UEs). We assume that time is slotted into discrete time intervals indexed by $t \in \mathbb{N}$. The time slot duration $\Delta t$ (seconds) is greater than the amount of time required to determine if a relay transmission is needed, identify an appropriate relay, and complete a relay transmission, but not so long that the UEs' geographic locations or channel conditions change significantly during the time slot. In an LTE network, for example, the time slot duration could be set to the length of one *frame* (10 ms).

In any given time slot $t$, some UEs are scheduled to receive data from the BS on the downlink; however, some of these UEs may experience bad channel conditions due to pathloss, multipath effects, and shadowing, which will limit their downlink data rates. In this situation, intermediate UEs can act as relays to help deliver data from the BS to the destination UEs. To enable device-to-device (D2D) relaying, we exploit D2D communications technology [9][10][11]. Note that D2D relays are not the same as infrastructure relay nodes, which are included as a part of the current LTE-Advanced standard [5] and do not require incentives to cooperate.

For illustration, we consider a two-hop amplify-and-forward relaying strategy using a single intermediate relay (i.e., a transmission from the BS to the relay followed by a transmission from the relay to the destination). Additionally, we use a simple *mode and relay selection* strategy for deciding when a UE will request a relay and which relay it will select. Importantly, the strategies considered in this report are for illustration: the token system and optimization/learning framework proposed in Sections III and IV can work with any other relaying, mode, and relay selection strategies without significant modification.[2]

Before we present our illustrative mode and relay selection strategy in Section II.A.2, we provide expressions for the achievable transmission rates with and without relay transmission in Section II.A.1.

*1. Direct and Relay Assisted Transmission Rates*

Let $\Gamma_t^{ij}$ denote the received Signal-to-Interference plus Noise Ratio (SINR) on the link between node $i$ and UE $j$ in time slot $t$, where node $i$ could be the BS or another UE: i.e.,

$$\Gamma_t^{ij} = \frac{g_t^{ij} P_t^{ij}}{W_t^{ij} N_0 + I_t^j} , \qquad (1)$$

where $P_t^{ij}$ (W) is the transmission power on link $i \rightarrow j$, $g_t^{ij}$ is the channel gain, $W_t^{ij}$ (Hz) is the link's bandwidth, $N_0$ (W/Hz) is the noise power spectral density, and $I_t^j$ is the inter-cell interference experienced by UE $j$. Equipped with capacity achieving error control codes, the transmission rate on link $i \rightarrow j$ in time slot $t$ can be written as

$$r_t^{ij} = W_t^{ij} \log_2(1 + \Gamma_t^{ij}) \text{ (bits/s).} \qquad (2)$$

---

[2] Our framework can be applied without significant modification to any two-hop relaying scheme that uses a single intermediate relay (regardless of the relaying, mode, and relay selection strategies); however, the token system and optimization/learning framework would have to be adapted to work with either two-hop relaying strategies that use *multiple* relays or *multi-hop* relaying strategies.

Suppose that, in time slot $t$, UE $j \in \{1,...,N\}$ is scheduled to receive data from the BS (indexed by 0). UE $j$ can either receive data directly over the B2D link (*direct mode*) or through an intermediate relay (*relay mode*). In **direct mode**, the BS transmits to UE $j$ at the rate $r_t^{0j} = W_t^{0j} \log_2(1 + \Gamma_t^{0j})$ (bits/s) for $\Delta t$ seconds. In **relay mode**, following standard relay channel analysis for amplify-and-forward cooperation [8], we obtain the following received SINR over the cooperative link:

$$\Gamma_t^{0ij} = \frac{\Gamma_t^{0i} \Gamma_t^{ij}}{\Gamma_t^{0i} + \Gamma_t^{ij} + 1}, \tag{3}$$

such that the effective transmission rate from the BS to UE $j$, through UE $i$, is $r_t^{0ij} = W_t^{0j} \log_2(1 + \Gamma_t^{0ij})$. Note that the relay-to-destination link uses the same physical resources as the BS-to-relay link. We discuss this in more detail in the next subsection.

*2. Illustrative Mode Selection and Relay Selection Strategies*

In practice, mode selection, relay selection, and scheduling should be optimized jointly in order to best utilize the available downlink resources; however, for simplicity, we assume that the BS first schedules the UEs for downlink transmission, and then each scheduled UE determines its transmission mode and, if necessary, its desired relay. Furthermore, we assume that the scheduler operates under the assumption that all transmissions are performed in the direct mode. In this way, we can integrate D2D relaying into the system without requiring any modification to the scheduler.

Let $\mathcal{U}_t^{DL}$ and $\mathcal{U}_t^{UL}$ denote the sets of UEs that are scheduled on the downlink and uplink, respectively, and let $\mathcal{U}_t^{IDLE}$ denote the set of idle UEs in time slot $t$. Note that, since UEs cannot transmit and receive at the same time, $\mathcal{U}_t^{DL}$, $\mathcal{U}_t^{UL}$, and $\mathcal{U}_t^{IDLE}$ partition the set of UEs $\mathcal{U} = \{1,...,N\}$. The sets of scheduled users, as well as their corresponding resource allocations, can be determined using any scheduling policy (proportional fair, round robin, resource fair, etc.). We let $\mathcal{U}_t^{IDLE}(j)$ denote the set of idle UEs that are associated with the same BS as UE $j$. Only UEs in $\mathcal{U}_t^{IDLE}(j)$ can serve as relays for UE $j$ in time slot $t$.

Consider UE $j \in \mathcal{U}_t^{DL}$, which is scheduled to receive data from the BS in time slot $t$ over a link with bandwidth $W_t^{0j}$ Hz. We assume that, if the received SINR on the BS to device (B2D) link, $\Gamma_t^{0j}$, is above a target threshold $\Gamma^{\text{target}}$, then the BS will transmit to UE $j$ in the direct mode. Note that, given the allocated bandwidth $W_t^{0j}$, the target SINR corresponds to a target rate $r^{\text{target}} = W_t^{0j} \log_2(1 + \Gamma^{\text{target}})$. However, if the B2D link's SINR is below the target threshold, then UE $j$ will attempt to find a relay $i^* \in \mathcal{U}_t^{IDLE}(j)$ through which it can meet the target rate while using the least power. UE $j$ achieves this by solving the following *mode and relay selection* problem:

$$i^* = \arg\min_{i \in \mathcal{U}_t^{IDLE}(j)} P_t^{ij}$$

s.t.

$$\begin{aligned}
\text{Rate constraint:} \quad & r_t^{0ij} \geq W_t^{0j} \log_2(1 + \Gamma^{\text{target}}) \\
\text{SNR constraint:} \quad & \Gamma_t^{0j} < \Gamma^{\text{target}} \\
\text{Power constraint:} \quad & P_t^{ij} \leq P^{\max} \\
\text{BW constraints:} \quad & W_t^{0i} = W_t^{0j} \text{ and } W_t^{ij} = W_t^{0j}.
\end{aligned} \tag{4}$$

If there is a relay $i^*$ that is a solution to problem (4), and is willing to provide relay services, then UE $j$ will receive data from the BS, through UE $i^*$, in relay mode; otherwise, UE $j$ will receive data from the BS in direct mode.

In problem (4), the rate constraint requires that the BS can meet the target transmission rate through the relay; the SINR constraint requires that the B2D link's SINR is below the target threshold; the power constraint requires that the relay's transmission power is below its maximum transmission power; and the bandwidth (BW) constraints require that the BS-to-relay and relay-to-destination links use the same resources that the scheduling policy originally allocated to the B2D link. Importantly, since the relay transmission is performed in two phases, the BW constraint $W_t^{ij} = W_t^{0j}$ implies that the D2D link is given dedicated (orthogonal) resources; therefore, we do not have to worry about the D2D communications causing interference to the BS or other UEs in the cell. This is in contrast to a lot of work on D2D communications, which assumes that D2D links reuse uplink resources, e.g., [10]. In the case of downlink relaying, we believe that it makes more sense for the D2D relays to reuse downlink resources because they are helping the base station.

*B. Virtual Token Exchange System*

Performing a relay transmission costs energy and provides no benefit to the relay. Consequently, without proper incentives, no self-interested user would want her UE to relay data for other users. To overcome this problem, we incentivize relaying through the use of a *virtual token exchange system.*

In our proposed system, UEs are initially endowed with a set of tokens (e.g., by the device's manufacturer or by the network operator). A UE must "expend" one token in order to receive data through a relay, and a UE can only "earn" additional tokens by relaying data for other UEs. Importantly, if a UE does not have any tokens, then it cannot receive data through a relay, so it will default to receiving data directly from the BS. In this way, no user is worse off in the proposed system than they would be in a network with only B2D communications.

Consider UE $j \in \mathcal{U}_t^{DL}$, which is scheduled to receive data from the BS in time slot $t$. Suppose that $\Gamma_t^{0j} < \Gamma^{\text{target}}$ and that, after solving problem (4), it selects UE $i^*$ as a candidate relay. At this point, UE $j$ sends a *relay request* to UE $i^*$, which replies with either a positive or negative *relay acknowledgement* (R-ACK/R-NACK) indicating if it is willing to relay or not (in Sections III and IV, we describe how UE $i^*$ makes this decision). If UE $j$'s request is R-ACK'd, then it receives data from the BS, through UE $i^*$, in relay mode, and a virtual token transfer takes place (from UE $j$ to UE $i^*$); otherwise, UE $j$ receives data directly from the BS.

Similar to [24], we assume that each UE is equipped with a secure and tamper proof hardware module, which is located between the UE's data link and physical layers. The secure hardware module keeps track of how many electronic tokens the UE currently holds and helps manage the virtual token exchange process.[3] We say that the token exchange is *virtual* because tokens are not actually exchanged between UEs; instead, upon receiving data through a relay, a UE's token holding is automatically decreased by one, and upon completing a relay transmission, a UE's token holding is automatically incremented by one. Similar to [22], in order to prevent malicious users from cheating the token system for their own benefit, we assume that the BS uses public-key cryptography to create *digital signatures*, which the relay and destination UEs can use to verify that messages were in fact generated by the BS and, furthermore, verify that the messages were not tampered with.[4]

Our proposed virtual token exchange system works as follows:

1. All base stations owned by a network operator share a common private key and an associated public key. All UEs know the public key. This can be achieved by having the device's manufacturer or the network operator load the public key on each UE before it is sold to the end user. A benefit of this approach is that the public key is very unlikely to be tampered with by a malicious third party and, therefore, we can be confident that it is authentic.
2. Before a base station sends data through a relay, it uses the private key to digitally sign the data as shown in Fig. 1(a).
3. The relay UE's secure hardware module uses the public key to authenticate the data it transmits (i.e., determine that it was generated by the BS) as shown in Fig. 1(b). If the data is authentic, then the relay's hardware module increments its token holding by one. Since the secure hardware module is located between the data link and physical layers, it will be very challenging for a malicious user to modify the data after it has been authenticated; therefore, this system prevents malicious UEs from earning tokens if they tamper with the data before relaying it.

---

[3] Note that, since UEs will already have to be designed to support D2D relaying, integrating the secure hardware module will require very little additional engineering and cost overheads. Moreover, since consumers frequently upgrade their mobile handsets, UEs with these capabilities can be widely deployed in a reasonably short time frame (e.g. 1-2 years).

[4] Digital signatures are analogous to sealing a message in an envelope with a personal wax seal: the envelope can be opened by anyone, but the unique seal authenticates the sender; moreover, if the seal is broken, then the receiver knows that the message inside the envelope may have been tampered with.

4. The destination UE's secure hardware module uses the public key to authenticate the data it receives and verify its integrity (i.e., determine that it was not altered by the relay) as shown in Fig. 1(c). If the data is authentic and has not been tampered with, then the destination UE's hardware module decrements its token holding by one. This prevents destination UEs from losing tokens if they receive data that has been tampered with.

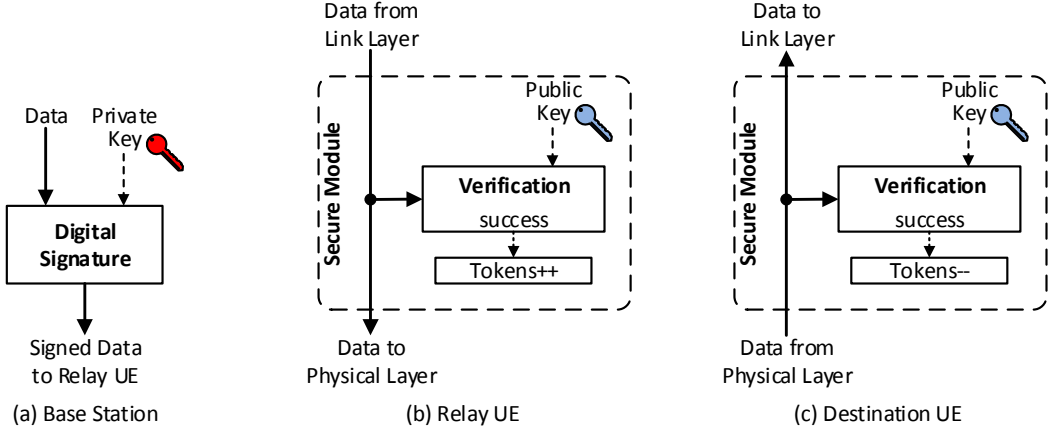

Fig. 1. Virtual token exchange system architecture. (a) Digital signature generation at the BS. (b) Data verification in the relay UE's transmit path. (c) Data verification in the destination UE's receive path.

*C. UE Model*

We now describe our model of UE $j \in \{1,...,N\}$.

**Token holding state:** At any given time, UE $j$ holds $k^j \in \mathcal{K} = \{0,1,...,T\}$ tokens, where $T$ is the total number of tokens in the network. As described in Section II.B, a UE must "expend" one token in order to receive data through a relay, a UE can only "earn" additional tokens by relaying data for other UEs, and, if a UE does not have any tokens, then it cannot receive data through a relay.

**Relay energy budget state:** Since UEs are mobile, they have limited battery energy. We let $p_{\max}^j$ represent the total amount of battery energy that the user is willing to consume relaying data for other users over the course of one battery charge. $p_{\max}^j$ can be set based on user preferences. With each relay transmission that UE $j$ provides, it expends some energy, and therefore reduces its relay energy budget $p^j \in [0, p_{\max}^j]$. When its relay energy budget reaches 0, UE $j$ will no longer relay data for other UEs and can no longer receive relay service from other UEs; however, it will continue to receive data over the direct B2D link. In other words, $p^j = 0$ does not mean that the UE's battery is completely drained; it only means that the UE will no longer participate in the token system. We refer to $p^j = 0$ as the *dead state* and assume that $p^j$ is reset to $p_{\max}^j$ when the UE's battery is recharged. Importantly, the UE keeps any tokens that it holds when it enters the dead state and can use them once its battery is recharged. Note that, in our problem formulation, we assume that $p^j$ is continuous (see Section III); however, as in [34], we quantize $p^j$ in our simulation results to make our proposed optimization problem tractable.

The three parameters described in the following three paragraphs are used to characterize the network environment experienced by UE $j$. These parameters are *unknown* a priori and *time-varying* because, as summarized in Table 1, they depend on various environmental factors including UE $j$'s geographic location, its distance from the nearest BS, the locations of other UEs in the network, the channel conditions, and the cooperation strategies, token holdings, and relay energy budgets of the other UEs. The definitions of each of these parameters are illustrated in Fig. 2.

Table 1. Parameters that characterize UE $j$'s experienced network environment, and their dependence on various environmental factors (✓ indicates dependence).

| **Environmental Factors** | **Environmental Parameters** | | |
|---|---|---|---|
| | ORDR $\lambda^j$ | IRDR $\mu^j$ | RRE $e^i$ |
| *Geographic location* | ✓ | ✓ | ✓ |
| *Distance to nearest BS* | ✓ | ✓ | ✓ |
| *Channel conditions* | ✓ | ✓ | ✓ |
| *Locations, cooperation strategies, token holdings, and relay energy budgets of other UEs* | ✗ | ✓ | ✓ |

**Outbound relay demand rate (ORDR):** We let $\lambda^j$ denote the probability that UE $j$ wants to get help from a relay to receive data from the BS. Given our illustrative mode and relay selection strategy [see Section II.A.2], in which a UE seeks a relay if its SINR is below a target threshold, we have

$$\lambda^j = \Pr\left(\Gamma_t^{0j} < \Gamma^{\text{target}}\right), \tag{5}$$

which means that $\lambda^j$ is equivalent to the outage probability of the B2D link to UE $j$. Importantly, $\lambda^j$ is the probability that UE $j$ *wants* help from a relay, but it is not the probability that it actually *receives* help from a relay. For this reason, we refer to $\lambda^j$ as UE $j$'s *outbound relay demand rate (ORDR)*.

**Inbound relay demand rate (IRDR):** We let $\mu^j$ denote the probability that UE $j$ is asked to relay data for another UE. We refer to $\mu^j$ as UE $j$'s inbound relay demand rate (IRDR).

**Relay recruitment efficiency (RRE):** Let $e^j$ denote the $j$ th UE's relay recruitment efficiency (RRE), which is defined as the following conditional probability:

$$e^j = \Pr(\text{R-ACK received}|\Gamma_t^{0j} < \Gamma^{\text{target}}, k^j > 0, p^j > 0). \tag{6}$$

In words, $e^j$ is the conditional probability that UE $j$ gets help from a relay (i.e., receives a R-ACK), given that it requires a relay transmission, has at least one token to "pay" a relay, and has a non-zero relay energy budget. The unconditional probability that UE $j$ receives a R-ACK can be written as $\lambda^j e^j I_{\{k^j > 0,\ p^j > 0\}}$, where $I_A$ is an indicator variable that is set to 1 when the event $A$ occurs and is set to 0 otherwise.

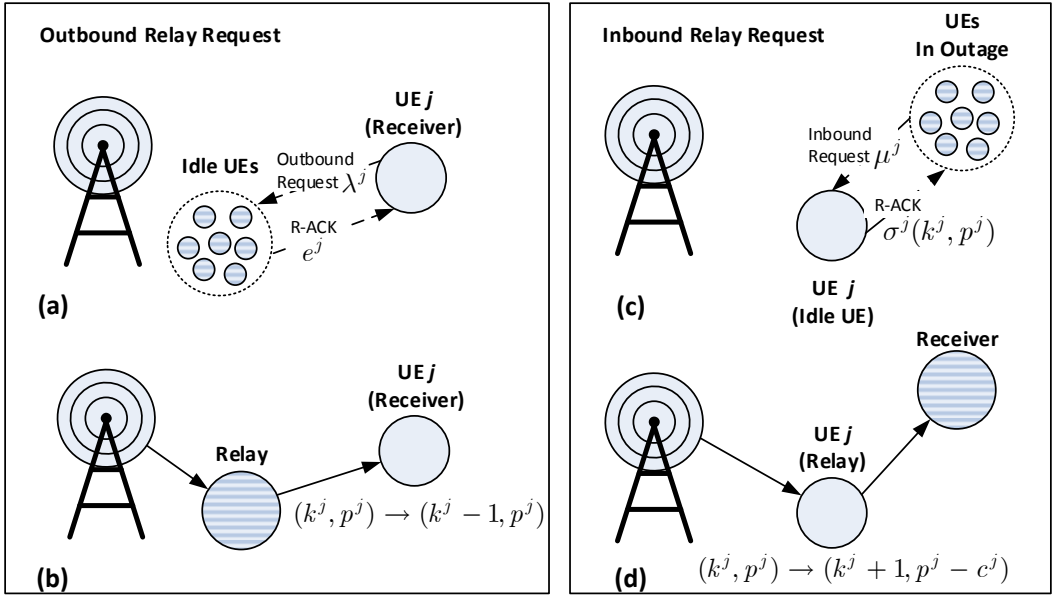

Fig. 2. Illustration of outbound and inbound relay requests in a network with cooperative downlink transmission assuming that $k^j > 0$ and $p^j > 0$. (a) UE $j$ seeks a relay from the pool of idle UEs with probability $\lambda^j$ and its request is R-ACK'd with probability $e^j$. (b) After its relay request is R-ACK'd, UE $j$ gets help from a relay in exchange for one token. (c) UE $j$ receives a relay request with probability $\mu^j$ and R-ACK's the request based on its cooperation action $\sigma^j(k^j, p^j)$. (d) After it R-ACK's the request, UE $j$ acts as a relay and expends $c^j$ units of energy in exchange for one token.

**Cost and benefit:** We let $b^j$ be the benefit gained by UE $j$ when it receives data through a relay and let $c^j$ be the energy cost incurred by UE $j$ when it provides a relay transmission to another UE. We assume that $b^j = \alpha \log_2(1 + \Gamma^{\text{target}})$ bits/s/Hz, which is fixed for all relay transmissions, and that $c^j = \Delta t P^{ji}$ Joules, where $P^{ji}$ is the transmission power used by UE $j$ to relay data to UE $i$ [recall from (4) that the selected relay can meet the target rate using the least transmission power]. The constant $\alpha$ is set such that $b^j > c^j$; if this condition does not hold, then UE's have no incentive to cooperate [30].

**Cooperation actions:** When a UE receives a relay request, it must decide if it will R-ACK the request. We define the action set $\mathcal{A}(p^j)$ as a function of the UE's relay energy budget $p^j$: i.e.,

$$\mathcal{A}(p^j) = \begin{cases} \{0,1\}, & p^j > 0 \\ \{0\}, & p^j = 0, \end{cases} \qquad (7)$$

where $a^j = 1 \in \mathcal{A}(p^j)$ (R-ACK) means that the UE is willing to act as a relay and $a^j = 0 \in \mathcal{A}(p^j)$ (R-NACK) means that it is not.[5] If the UE is in the dead state, then it will not act as a relay.

---

[5] One may argue that $\mathcal{A}(p^j) = \{0,1\}$ only if $p^j > c^j$, where $c^j$ is the amount of energy required to relay. However, since $p^j$ is the relay energy budget and not the actual battery state, and $c^j \ll p^j_{\max}$, we allow relay transmissions as long as $p^j > 0$. This simplifies the model.

**Cooperation policy:** UE $j$'s cooperation policy $\sigma^j(k^j, p^j) \in \mathcal{A}(p^j)$ is a function, which maps its current token holding state $k^j$ and relay energy budget $p^j$ to a cooperation action $a^j \in \mathcal{A}(p^j)$. Given its IRDR, token holding state, relay energy budget, and cooperation policy, the probability that UE $j$ relays data for another UE in a time slot is $\mu^j \sigma^j(k^j, p^j) I_{\{p^j > 0\}}$. The role of the cooperation policy is illustrated in Fig. 2(c). Note that, a UE will only be able to get relay service as often as it provides relay service because it must earn as many tokens as it spends. We discuss this in more detail in Section III.B.

**State evolution:** We denote the UE's state by $s^j = (k^j, p^j) \in \mathcal{S}$, where $k^j \in \mathcal{K} = \{0,1,...,T\}$ is its token holding state and $p^j \in [0, p^j_{\max}]$ is its relay energy budget. When the UE acts as a relay, it gains one token and its relay energy budget reduces by $c^j$: i.e.,

$$\textbf{Act as a relay: } (k^j, p^j) \rightarrow (k^j + 1, p^j - c^j). \tag{8}$$

When the UE uses a relay, it loses one token and its relay energy budget remains the same: i.e.,

$$\textbf{Use a relay: } (k^j, p^j) \rightarrow (k^j - 1, p^j). \tag{9}$$

**Transition probability function:** We let $P^j([k^{j\prime}, p^{j\prime}] \mid [k^j, p^j], a^j)$ denote the state transition probability function, which gives the probability that UE $j$ transitions from state $s^j = (k^j, p^j)$ to state $s^{j\prime} = (k^{j\prime}, p^{j\prime})$ after taking cooperation action $a^j$. Based on (8) and (9), as well as the definitions of the ORDR $\lambda^j$, IRDR $\mu^j$, and RRE $e^j$, the state transition probability function is defined as follows:

$$P^j([k^{j\prime}, p^{j\prime}] \mid [k^j, p^j], a^j) = \begin{cases} \lambda^j e^j I_{\{k^j > 0, \, p^j > 0\}}, & \text{if } k^{j\prime} = k^j - 1 \text{ and } p^{j\prime} = p^j \\ \mu^j a^j I_{\{p^j > 0\}}, & \text{if } k^{j\prime} = k^j + 1 \text{ and } p^{j\prime} = p^j - c^j \\ 1 - \lambda^j e^j I_{\{k^j > 0, \, p^j > 0\}} - \mu^j a^j I_{\{p^j > 0\}}, & \text{if } k^{j\prime} = k^j \text{ and } p^{j\prime} = p^j \\ 0, & \text{otherwise,} \end{cases} \tag{10}$$

Intuitively, in each time slot, a UE can provide help as a relay, get help from a relay, or do neither: if UE $j$ has non-zero tokens and a non-zero relay energy budget (i.e., $I_{\{k^j > 0, \, p^j > 0\}} = 1$), then with probability $\lambda^j e^j$ it gets help and pays one token [line 1 in (10)]; if UE $j$ has a non-zero relay energy budget (i.e., $I_{\{p^j > 0\}} = 1$), then with probability $\mu^j a^j$ it provides help, gains one token, and loses $c^j$ units of battery energy [line 2 in (10)]; if UE $j$ has non-zero tokens and a non-zero relay energy budget, then with probability $1 - \lambda^j e^j - \mu^j a^j$ it neither gets help nor provides help [line 3 in (10)]; if UE $j$ has zero tokens or is in the dead state (i.e., $I_{\{k^j > 0, \, p^j > 0\}} = 0$), then it cannot receive help; if UE $j$ is in the dead state (i.e., $I_{\{p^j > 0\}} = 0$), then it cannot provide help; and, all other cases occur with probability 0.

**Expected utility:** Let $u^j(k^j, p^j, a^j)$ denote UE $j$'s expected utility in state $s^j = (k^j, p^j)$ when it takes cooperation action $a^j$. The expected utility is defined as follows:

$$u^j(k^j, p^j, a^j) = I_{\{k^j > 0, \, p^j > 0\}} \lambda^j e^j b^j - I_{\{p^j > 0\}} \mu^j a^j c^j. \tag{11}$$

Intuitively, if UE $j$ has non-zero tokens and a non-zero relay energy budget, then it gets help from a relay with probability $\lambda^j e^j$ and receives benefit $b^j$; if UE $j$ has a non-zero relay energy budget, then it provides help with probability $\mu^j a^j$ and incurs cost $c^j$; and, if UE $j$ has a zero relay energy budget, then it can neither provide help nor receive help, therefore, its utility within the token system is 0.

## III. OPTIMAL COOPERATION POLICY FOR A SINGLE UE

### A. Problem Formulation

In this section, we formulate the problem of determining a UE's optimal cooperation policy as an MDP [31]. We assume that its ORDR $\lambda^j$, IRDR $\mu^j$, and RRE $e^j$ are known and fixed, and therefore its transition probability and utility functions are known and fixed; however, in Section IV, we consider the case when these parameters are unknown a priori and time-varying, and propose a learning algorithm that each UE can deploy to dynamically adapt its cooperation policy to its experienced network environment.

Let $u_t^j$ denote UE $j$'s utility in time slot $t$ and let $\mathbf{E}[u_t^j] = u^j(k_t^j, p_t^j, a_t^j)$ denote its expectation. Each UE aims to maximize its *infinite horizon discounted utility*[6] in which benefits and costs that are received $t$ time steps in the future are discounted by the factor $\beta^t$, where $\beta \in [0,1)$: that is,

$$\mathbf{E}\left[\sum_{t=0}^{\infty} \beta^t u_t^j\right]. \tag{12}$$

We assume that the discount factor $\beta$ is a common parameter among all UEs and that it is set by the network operator. If $\beta = 0$, then each UE only considers its immediate utility, and will therefore never choose to act as a relay (because it will incur cost $c^j$ without any immediate benefit to itself). In general, larger values of $\beta$ compel each UE to look farther into the future to determine its optimal cooperation action at the present time. If the network environment is stationary, then discount factors closer to 1 will always lead to better long run performance; however, since the ORDR $\lambda^j$, IRDR $\mu^j$, and RRE $e^j$ are time-varying in practice, forecasts of the future may be inaccurate, so discount factors too close to 1 may

---

[6] There are other optimization objectives that one might think to use; however, they are not well suited for the problem considered in this report. For example, a finite horizon undiscounted utility objective [31], i.e., $\mathbf{E}\left[\sum_{t=0}^{h} u_t^j\right]$, results in a non-stationary cooperation policy, which requires a large amount of memory to store due to the long time horizons that are required in our optimization. This is further exacerbated by the fact that the learning solution we propose in Section IV requires pre-computing and storing many cooperation strategies. Alternatively, we could use receding-horizon control [32]. However, due to the long time horizons required in our optimization, computing the optimal decision in each time step would be computationally infeasible.

As another example, an infinite horizon undiscounted utility objective [31], i.e., $\lim_{h \to \infty} \frac{1}{h} \mathbf{E}\left[\sum_{t=0}^{h} u_t^j\right]$, cannot be used for our problem because the Markov chain induced by any non-trivial cooperation policy (in which the UE chooses to cooperate) is non-ergodic. This is due to the fact that there is an absorbing state in the MDP.

We have chosen to use an infinite-horizon discounted formulation because it is guaranteed to have a stationary optimal policy regardless of the structure of the MDP, and it is less computationally complex than the finite-horizon case for long time horizons (corresponding to discount factors close to 1).

actually lead to worse long run performance. Determining the optimal discount factor is out of the scope of this report, but we will consider the problem in future research.

We define UE $j$'s optimal *state value function* $V^j(k^j, p^j)$ as the expected infinite horizon discounted utility that it will gain from each state if it executes the optimal policy $\sigma^j(k^j, p^j)$: that is,

$$V^j(k^j, p^j) = \max_{\sigma^j} \mathbf{E}\left[\sum_{t=0}^{\infty} \beta^t u_t^j\right] \tag{13}$$

The optimal state value function is unique and satisfies the following Bellman optimality equation:

$$V^j(k^j, p^j) = \max_{a^j \in \mathcal{A}(p^j)} \underbrace{\left\{u^j\left(k^j, p^j, a^j\right) + \beta \sum_{(k^{j'}, p^{j'}) \in \mathcal{S}} P([k^{j'}, p^{j'}] \mid [k^j, p^j], a^j) V(k^{j'}, p^{j'})\right\}}_{Q^j\left(k^j, p^j, a^j\right)}. \tag{14}$$

In (14), we also defined the *action-value function* $Q^j(k^j, p^j, a^j)$, which is the value of taking cooperation decision $a^j$ in state $s^j = (k^j, p^j)$ and then following the optimal policy thereafter. Given the action-value function, it is easy to determine the optimal cooperation policy:

$$\sigma^j(k^j, p^j) = \arg\max_{a^j \in \mathcal{A}(p^j)} Q^j(k^j, p^j, a^j). \tag{15}$$

Importantly, the optimal value function can be computed using the well-known value iteration algorithm:

$$V_{n+1}^j(k^j, p^j) = \max_{a^j \in \mathcal{A}(p^j)} \left\{u^j(k^j, p^j, a^j) + \beta \sum_{(k^{j'}, p^{j'}) \in \mathcal{S}} P([k^{j'}, p^{j'}] \mid [k^j, p^j], a^j) V_n^j(k^{j'}, p^{j'})\right\}, \tag{16}$$

where $V_0^j(k^j, p^j) = 0$ for all $s^j = (k^j, p^j) \in \mathcal{S}$ and $V_n^j(k^j, p^j)$ converges to the optimal state-value function $V^j(k^j, p^j)$ as $n \to \infty$ [31].

In practice, the cost and transition probability functions are unknown a priori and time-varying (due to the fact that the ORDR $\lambda^j$, IRDR $\mu^j$, and RRE $e^j$ are unknown and time-varying), so UE $j$ cannot directly apply value iteration to find the optimal policy; instead, it must learn its optimal cooperation policy online based on its experience. We propose a solution to the online problem in Section IV.

*B. Balance of Earned and Expended Tokens*

We noted in Section II.C that, on average, a UE will only be able to get relay service as often as it provides relay service because it must earn as many tokens as it spends. In this section, we characterize this balance under some simplifying assumptions. Let $k_t^{j,+} \geq 0$ and $k_t^{j,-} \geq 0$ denote the cumulative number of tokens earned and expended by UE $j$ from time slot 0 to time slot $t$, respectively, and let $k_0^j \geq 0$ denote its initial token allocation. Within the proposed token system, the following condition must always hold:

$$k_t^{j,-} \leq k_t^{j,+} + k_0^t, \tag{17}$$

which indicates that the number of expended tokens cannot exceed the number of earned tokens plus the number of initial tokens. Suppose that UE $j$ has an unlimited relay energy budget (and an unlimited battery capacity), and that its ORDR $\lambda^j$, IRDR $\mu^j$, and RRE $e_j$ are fixed so that we can study the steady-state behavior of (17).[7] If we divide both sides of (17) by $t$ and take the limit as $t \to \infty$, then we get

$$\underbrace{\Pr(\text{UE } j \text{ receives help from a relay})}_{-1 \text{ Token}} \leq \underbrace{\Pr(\text{UE } j \text{ provides help as a relay})}_{+1 \text{ Token}}, \qquad (18)$$

because UE $j$ expends one token every time that it receives help from a relay and it earns one token every time that it provides help as a relay. The condition in (18) can be rewritten as:

$$\underbrace{\lambda^j e^j \mathbf{E}_{k^j}[I_{\{k^j > 0\}}]}_{-1 \text{ Token}} \leq \underbrace{\mu^j \mathbf{E}_{k^j}[\sigma^j(k^j, p^j)]}_{+1 \text{ Token}}, \qquad (19)$$

where the left hand side is the probability that UE $j$ uses a relay (and expends one token), the right hand side is the probability that UE $j$ acts as a relay (and earns one token), $\mathbf{E}_{k^j}[\cdot]$ denotes the expectation over the UE's steady-state token holding distribution, $\mathbf{E}_{k^j}[I_{\{k^j > 0\}}]$ is the probability that UE $j$ has non-zero tokens, and $\mathbf{E}_{k^j}[\sigma^j(k^j, p^j)]$ is the probability that UE $j$ will R-ACK an inbound relay request. Note that equality holds in (19) if $\sigma^j(k^j, p^j)$ is the optimal policy; if equality does not hold, then UE $j$ earns more tokens than it can spend, and therefore consumes more energy relaying than necessary.

*C. Threshold Strategies*

UE $j$'s cooperation decision depends on its token holding and relay energy budget states. Our simulation results in Section V.B show that the optimal cooperation policy $\sigma^j(k^j, p^j)$ is threshold in the token state $k^j$ and that the threshold depends on the relay energy budget $p^j$: specifically,

$$\sigma^j(k^j, p^j) = \begin{cases} 1, & \text{if } k^j \leq K_{\text{th}}(p^j), \\ 0, & \text{otherwise}, \end{cases} \qquad (20)$$

where the threshold $K_{\text{th}}(p^j)$ is non-decreasing in $p^j$. In other words, as a UE's relay energy budget drains, it will not want to earn as many tokens because it will have less opportunities to use them before it enters the dead state; therefore, it uses a lower threshold. A rigorous proof of this result is left as future work; however, in [28], we show that policies are threshold if UEs have unlimited relay energy budgets.

## IV. LEARNING THE OPTIMAL COOPERATION POLICY

Each UE experiences different environmental dynamics depending on its geographic location, its distance from the nearest BS, its channel conditions, and, importantly, the locations, cooperation strategies, token holdings, and relay energy budgets of the other UEs. The dynamics experienced by UE $j$ are captured by three parameters, namely, the ORDR $\lambda^j$, the IRDR $\mu^j$, and the RRE $e^j$, which are all defined in Section II.C. In this section, we propose a simple, low-complexity, supervised learning algorithm that a UE can deploy to learn its optimal cooperation policy $\sigma^j(k^j, p^j)$ online, despite the fact that these three parameters are unknown and time-varying.[8]

---

[7] If a UE's relay energy budget is finite, then there will be no token exchanges in steady-state. This is because, in steady-state, the UE's relay energy budget will be zero. By assuming that the relay energy budget and battery lifetime are both unlimited, we can use the steady-state behavior to approximate the behavior we observe while the UE's relay energy budget is non-zero, but finite.

Our proposed learning algorithm comprises an offline phase and an online phase. In the online phase, UE $j$ estimates its ORDR $\lambda^j$, IRDR $\mu^j$, and RRE $e^j$. Then, in each time slot, UE $j$ selects its cooperation policy based on the estimated values. In principle, UE $j$ could use these estimated values to populate the transition probability and utility functions defined in (10) and (11), respectively, and then use value iteration [see (16)] to compute the corresponding optimal cooperation policy; however, recomputing the optimal policy in every time slot would be computationally prohibitive.

For this reason, we propose to first compute a collection of cooperation policies *offline*, which correspond to a representative set of discretized environmental parameters, and then use a simple look-up table to select UE $j$'s cooperation policy *online* in each time slot. To reduce the size of the look-up table, instead of estimating the ORDR $\lambda^j$ and RRE $e^j$ independently, we directly estimate $\pi^j = \lambda^j e^j \in [0, 1/2]$, which we refer to as the *outbound relay success rate*. Note that the maximum outbound relay success rate is ½ because, on average, a UE can only receive relay service as often as it provides relay service (see Section III.B). Additionally, since the optimal policy is threshold in the token state (see Section III.C), each policy can be represented compactly with only one (threshold) value per relay energy budget state. We provide pseudo-code for the offline and online phases in Table 2 and Table 3, respectively, and describe each phase in more detail in Sections IV.A and IV.B, respectively.

*A. Offline Phase*

Let $\Pi = \{\pi^1, \pi^2, ..., \pi^X\}$, $\mathrm{M} = \{\mu^1, \mu^2, ..., \mu^Y\}$, and $\mathcal{C} = \{c^1, c^2, ..., c^Z\}$ be finite sets containing representative values of the outbound relay success rate, IRDR, and relay cost, respectively. In the offline phase of the proposed learning algorithm, we compute the collection of policies $\{\sigma(k, p \mid \pi, \mu, c) : \pi \in \Pi, \mu \in \mathrm{M}, c \in \mathcal{C}\}$. To compute the policy $\sigma(k, p \mid \pi, \mu, c)$, we first populate the transition probability and utility functions defined in (10) and (11), respectively, with $\pi$, $\mu$, and $c$ in place of $\lambda^j e^j$, $\mu^j$, and $c^j$ (lines 3-4 in Table 2). Subsequently, we use value iteration to compute the optimal state-value function and the optimal cooperation policy corresponding to the representative parameters (lines 5-6 in Table 2).

---

[8] We do not use a more conventional reinforcement learning based approach, such as a Q-learning [31], because our problem does not satisfy the required convergence conditions. Specifically, Q-learning requires that all state-action pairs are visited "infinitely often": that is, in the limit as $t \to \infty$, every action must be taken in every state an infinite number of times. This condition does not hold for our problem because the relay energy budget is non-increasing in time. We refer the interested reader to the previous version of this technical report [39] for an in depth study of the shortcomings of traditional reinforcement learning algorithms in the context of our problem.

Since the sets of possible outbound relay success rates, IRDRs, and relay costs are discretized into $X$, $Y$, and $Z$ values, respectively, a total of $XYZ$ policies must be computed offline and stored in a look-up table. Assuming that the relay energy budget is discretized into $W$ values, the total amount of memory required to store the look-up table is $O(WXYZ)$ (because each policy can be compactly represented with one threshold value per relay energy budget state). For example, if each variable is discretized into 10 values, and the threshold is represented by one byte, then the lookup table will require 10 KB of memory.

Importantly, the offline phase can be performed before the consumer purchases her cellular device and the collection of optimal policies can be preloaded onto the device (e.g., by the device's manufacturer or by the network operator). Additionally, the offline phase can be performed again whenever the consumer changes her relaying preferences, i.e., changes her maximum relay energy budget $p_{\max}^j$.

Table 2. Learning algorithm: Offline phase

| 1. | **Input:** $\Pi = \{\pi^1, \pi^2, ..., \pi^X\}$, $M = \{\mu^1, \mu^2, ..., \mu^Y\}$, $\mathcal{C} = \{c^1, c^2, ..., c^Z\}$, $p_{\max}$, and the number of quantized relay energy budget states |
|---|---|
| 2. | **For** each $(\pi, \mu, c) \in \Pi \times M \times \mathcal{C}$ |
| 3. | Populate the transition probability function $P([k', p'] \mid [k, p], a)$ as defined in (10) |
| 4. | Populate the utility function $u(k, p, a)$ as defined in (11) |
| 5. | Calculate optimal state-value function $V(k, p)$ using value iteration (16) |
| 6. | Calculate the optimal cooperation policy $\sigma(k, p)$ using (15) |
| 7. | Record the optimal cooperation policy $\sigma(k, p \mid \pi, \mu, c) \leftarrow \sigma(k, p)$ |
| 8. | **End** |
| 9. | **Output:** Collection of policies $\{\sigma(k, p \mid \pi, \mu, c) : \pi \in \Pi, \mu \in M, c \in \mathcal{C}\}$ |

*B. Online Phase*

In the online portion of the algorithm, each UE maintains online estimates of its outbound relay success rate $\pi^j$, denoted by $\hat{\pi}_t^j$, and its IRDR $\mu^j$, denoted by $\hat{\mu}_t^j$. These estimates can be determined using an exponential moving average of successful outbound relay requests and inbound relay requests, respectively (lines 11-12 in Table 3). Upon receiving an inbound relay request (line 6 in Table 3), UE $j$ evaluates the energy cost $c_t^j$ it will incur if it provides a relay transmission (line 8 in Table 3). UE $j$ then takes the action $a_t^j = \sigma^j(k_t^j, p_t^j \mid f(\hat{\pi}_t^j), g(\hat{\mu}_t^j), h(c_t^j))$, where $f : [0, 1/2] \to \Pi$ maps the estimate $\hat{\pi}_t^j$ to the nearest value in $\Pi$, $g : [0,1] \to M$ maps the estimate $\hat{\mu}_t^j$ to the nearest value in $M$, and $h : [0,1] \to \mathcal{C}$ maps the cost $c_t^j$ to the nearest value in $\mathcal{C}$ (line 9 in Table 3).

Table 3. Learning algorithm: Online phase.

| | |
|---|---|
| 1. | **Initialize:** $k_0^j$, $p_0^j = p_{\max}^j$, $\hat{\pi}_0^j$, $\hat{\mu}_0^j$, $f(\cdot)$, $g(\cdot)$, $h(\cdot)$, $w$ |
| 2. | **For** $t = 0, 1, \ldots$ |
| 3. | $outbound\_success \leftarrow 0$ and $inbound\_request \leftarrow 0$ |
| 4. | **If** $k_t^j > 0$ **and** $p_t^j > 0$ **and** UE $j$ has a successful outbound relay request |
| 5. | $outbound\_success \leftarrow 1$ |
| 6. | **Else if** $p_t^j > 0$ **and** UE $j$ receives an inbound relay request |
| 7. | $inbound\_request \leftarrow 1$ |
| 8. | Determine energy cost $c_t^j$ for relay transmission |
| 9. | Take cooperation action $a_t^j \leftarrow \sigma^j(k_t^j, p_t^j \mid f(\hat{\pi}_t^j), g(\hat{\mu}_t^j), h(c_t^j))$ |
| 10. | **End** |
| 11. | $\hat{\pi}_{t+1}^j \leftarrow \dfrac{1}{w} \times outbound\_success + \dfrac{w-1}{w} \times \hat{\pi}_t^j$ |
| 12. | $\hat{\mu}_{t+1}^j \leftarrow \dfrac{1}{w} \times inbound\_request + \dfrac{w-1}{w} \times \hat{\mu}_t^j$ |
| 13. | Determine the next state as describe in Section II.C |
| 14. | **End** |

## V. SIMULATION RESULTS

In this section, we present our simulation results. In Section V.A, we describe the simulation setup that is used in Sections V.C-V.F. In Section V.B, we present several numerical results to highlight the structure of the optimal cooperation policies. Our focus in Sections V.B is on the behavior of a single UE's optimal (best response) cooperation policy given the aggregate behavior of the other UEs in the network. For this reason, the results in Sections V.B are generated outside of the network simulation so that we have full control over the UE's environment. In Sections V.C-V.F, we simulate a large-scale cellular network in which all users simultaneously deploy the supervised learning algorithm proposed in Section IV. Specifically, in Section V.C, we highlight the key differences between the proposed approach to cooperation (in which users are self-interested and must be incentivized to cooperate using the token system) and the conventional approach to cooperation (in which users are obedient and are obliged to cooperate even if it is not in their self-interest) in terms of *individual* user performance and *overall* network performance. Then, in Sections V.D and V.E, we investigate how UEs' mobility and relay energy budgets impact their own performance, the performance of other UEs, and the overall network performance within the proposed framework. Finally, in Section V.F, we study the impact of the network-wide token supply on the overall network performance.

*A. Cellular Network Simulation Setup*

In our illustrative simulation setup, we assume that $N = 1500$ mobile transceivers are uniformly and randomly distributed in a 10 km x 10 km square area consisting of 100 cells with size 1 km x 1 km. There is one BS at the center of each cell. In each time slot, each UE moves to a nearby location according to a random waypoint mobility model that is commonly used in the simulation of mobile networks [35][36] and may need to receive data from the BS in the corresponding cell. We assume a time slot duration of $\Delta t = 5$ s, which implies that downlink resources are allocated every 5 s and that the channel is invariant over this allocated time. In practice, the time slot duration would be much shorter to both match the time-scale at which most cellular networks schedule resources (e.g., 10 ms) and to ensure channel coherence in each time slot; however, in order to observe the effect of mobility on system performance, we need to operate on a larger time scale. Note that this is not a limitation of our proposed framework, but is instead a limitation of our ability to simulate a large-scale system with a small time slot duration. In particular, using a 10 ms time slot duration to simulate, e.g., one hour of real-time system operation, requires 500 times more computer time than using a 5 second time slot duration.

We divide users/UEs into two mobility classes and two relay energy budget classes:

1. **High mobility users** move at speeds between 50 and 120 km/hour (similar to [37] which uses a maximum speed of 30 m/s = 108 km/hour in its experiments). For instance, users in motor vehicles are considered high mobility users. These users play different roles in the network over time because sometimes they will be far from a base station where they will have high ORDRs, and sometimes they will be closer to a base station where they will have high IRDRs.
2. **Low mobility users** move at speeds between 0 and 8 km per hour. For example, users in offices, restaurants, or on foot are considered to be low mobility users. Due to their limited mobility, these users typically do not switch roles over time and their ORDRs and IRDRs are relatively static.
3. **High relay energy budget UEs** can R-ACK many relay requests before entering the dead state.
4. **Low relay energy budget UEs** can R-ACK fewer relay requests before entering the dead state.

The relay energy budgets of UEs in the high and low relay energy budget classes are specified separately for each of our simulations.

For computational reasons (i.e., so we can compute the optimal policy with value iteration), we quantize the relay energy budget state into eleven bins (one representing the dead state) and we limit the maximum number of tokens that the UE can hold to 20.[9]

We consider path loss and shadow fading for the channel model such that [33],

$$P_{\text{rx}} = P_{\text{tx}} - PL(d_0) - 10\eta \log(d / d_0) - \chi, \qquad (21)$$

where $P_{\text{rx}}$ and $P_{\text{tx}}$ are the receive and transmit powers (in dB), respectively, $PL(d_0)$ is the path loss of the reference distance $d_0$, $d$ is the distance between the source and destination, $\eta$ is the path loss factor, and $\chi$ is a Gaussian distributed random variable representing the effect of shadow fading. We assume that the maximum transmission power of the BS and UEs is 15 dBm, the total system bandwidth is 50 MHz, a maximum of 10 MHz can be allocated to each downlink transmission, and the target data rate is 1 bit/s/Hz. If the target downlink data rate cannot be achieved, then the UE requests a relay transmission. Using the above parameter values, the average ORDR throughout the network is approximately $\lambda = 0.1$. All UEs adapt their cooperation policies using the supervised learning algorithm proposed in Section IV and all UEs use a policy look-up table that is computed using the representative outbound relay success rates, IRDRs, and costs that are given in Table 4. Finally, we assume that there are $T = 9000$ tokens in the network that are uniformly and randomly distributed among the UEs at the start of the simulation. We discuss the impact the token supply in Section V.F. The parameters used in our simulations are summarized in Table 4.

---

[9] Note that maximum token holding is set to be above the highest cooperation threshold so that it does not affect the optimal policy.

Table 4. Simulation parameters.

| Parameter | Notation | Value |
| --- | --- | --- |
| Number of UEs | $N$ | 1500 |
| Time slot duration $\Delta t$ | $\Delta t$ | 5 s |
| Maximum transmit power | $P^{\max}$ | 15 dBm (32 mW) |
| Total BW / Max. BW allocation | N/A | 50 / 10 MHz |
| Path loss factor | $\eta$ | 3 |
| Reference distance / pathloss | $d_0$ / $PL(d_0)$ | 100 m / -5 dB |
| Noise power spectral density | $N_0$ | -10 dBm |
| Std. dev. of Gaussian fading | $\sqrt{\mathrm{var}(\chi)}$ | 2 dB |
| Target SINR | $\Gamma^{\text{target}}$ | 0 dB |
| Token supply | $T$ | Variable |
| Discount factor | $\beta$ | 0.99 |
| Outbound relay success rate set | $\Pi$ | $\{0.05, 0.1, \ldots, 0.45\}$ |
| Inbound relay demand rate set | $M$ | $\{0.05, 0.1, \ldots, 0.45\}$ |
| Cost set | $\mathcal{C}$ | $\{0.025, 0.05, \ldots, 0.225\}$ J |
| Benefit | $b^j$ | 0.5 |
| Learning window size | $w$ | 50 |
| No. quantized relay energy budget states | N/A | 11 (including dead state) |
| Relay energy budget | $p^j_{\max}$ | Variable |
| User mobility | N/A | Variable |

Before discussing our results, we need to introduce two new definitions:

- **R-ACK rate:** The *R-ACK rate* is the probability that, upon receiving a relay request, a UE sends a R-ACK. Under the assumptions in Section III.B, the R-ACK rate is equivalent to $\mathbf{E}_{k^j}[\sigma^j(k^j, p^j)]$.

- **Throughput gain:** The *throughput gain* is the ratio of the actual throughput to the direct transmission throughput. Since the actual throughput is always greater than or equal to the direct throughput, the minimum throughput gain is 1. Note that a throughput gain of 1 is achieved if users are self-interested and the token system is not in place to incentivize cooperation.

*B. Structure of the Optimal Policy*

In this section, we present several figures to highlight the structure of the optimal cooperation policy. Throughout this section, we fix certain parameters to illustrate the behavior of the optimal cooperation policy as we vary other parameters; importantly, the general behavior of the optimal cooperation policy with respect to the variable parameters does not depend on the specific values of the fixed parameters. We specify the fixed parameters in each figure's caption or title. We assume that each UE has a relay energy budget that allows it to R-ACK 1000 relay requests on average before entering the dead state.

In Fig. 3, we illustrate the optimal cooperation policy $\sigma(k,p)$. As we discussed in Section III.C, the optimal policy is threshold in the token state $k$, and the threshold decreases as the relay energy budget $p$ decreases. All policies that we have observed have this threshold structure.

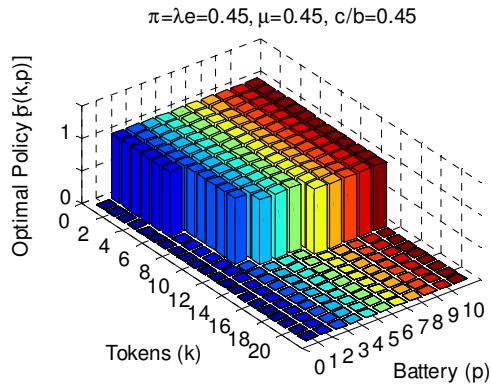

Fig. 3. Structure of the optimal cooperation policy $\sigma(k,p)$ with discount factor $\beta = 0.99$.

Fig. 4(a) illustrates how the optimal cooperation threshold varies with respect to the discount factor $\beta$ and the cost-to-benefit ratio $c/b$ (note that the policy depends on the cost-to-benefit ratio, rather than the absolute values of the costs and benefits [30]). The optimal cooperation threshold is non-decreasing in $\beta$ because, if a UE looks farther ahead, then it anticipates more opportunities to use additional tokens, and therefore has more incentive to cooperate in order to earn more tokens. Additionally, as we proved in our prior work [28], the optimal threshold decreases as the cost-to-benefit ratio $c/b$ increases.

Fig. 4(b) illustrates how the optimal decision threshold varies with respect to the cost-to-benefit ratio $c/b$ and IRDR $\mu$. The optimal decision threshold increases as the IRDR decreases because, if a UE is asked to relay infrequently, then it has incentive to earn tokens whenever it has the opportunity.

Fig. 4(c) illustrates how the optimal decision threshold varies with respect to the cost-to-benefit ratio $c/b$ and the outbound relay success rate $\pi = \lambda e$. We observe that the optimal decision threshold decreases as the outbound relay success rate decreases. This happens because, if a UE is unable to successfully recruit relays, then it does not have incentive to gather many tokens.

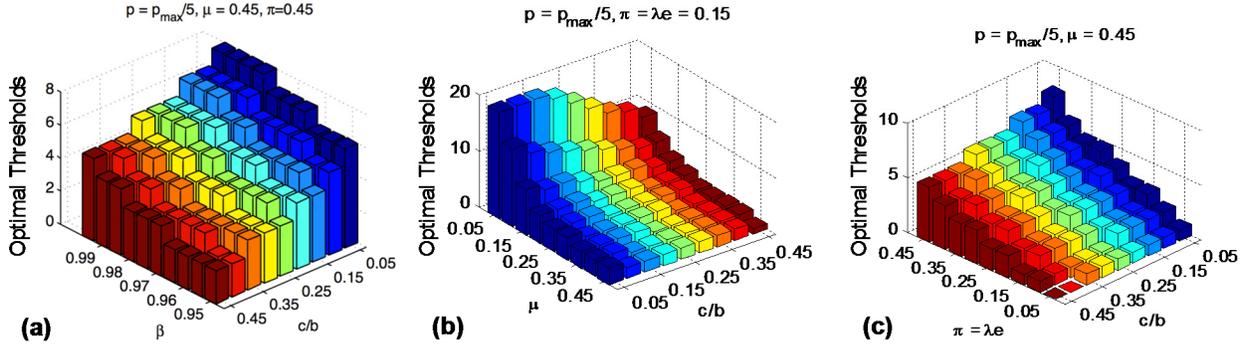

Fig. 4. Structure of the optimal cooperation policy. (a) Optimal thresholds for various values of $\beta$ and $c/b$. (b) Optimal thresholds for various values of $\mu$ and $c/b$. (c) Optimal thresholds for various values of $\pi$ and $c/b$. The results in (b) and (c) use discount factor $\beta = 0.99$.

## C. Comparison to Conventional Cooperative Relaying

The foundational assumption in this report is that UEs are *self-interested*. In other words, they aim to maximize their own utilities and do not care about the overall performance of the network. Since relaying costs energy and provides no immediate benefit to the relay, *self-interested UEs will never cooperate unless they are incentivized to do so*. Consequently, *without the proposed token-based incentives, self-interested UEs will never achieve more than the direct B2D throughput*. In contrast, the typical assumption in the cooperative communications literature (see, e.g., [2]-[4][7]-[11]) is that UEs are *obedient*, i.e., they will *always* cooperate even when it is not in their self-interest. To highlight the key differences between a network of self-interested users and a network of obedient users in terms of *individual* user performance and *overall* network performance, we compare the following three scenarios:

1. All UEs are self-interested, have finite relay energy budgets, and are incentivized to cooperate by using tokens (i.e., the proposed solution).
2. All UEs are obedient and have unlimited relay energy budgets (and unlimited battery energy).
3. All UEs are obedient and have limited relay energy budgets.

We use the simulation test-bed described in Section V.A. We assume that 70% (30%) of users have high (low) mobility and that 70% (30%) of users have high (low) relay energy budgets. We assume that UEs with high (low) relay budgets can relay 100 (40) times on average before entering the dead state. These choices of high and low relay energy budgets allow us to highlight how the network performance changes as increasing numbers of UEs with low relay energy budgets transition to the dead state while UEs with high relay energy budgets remain active. Importantly, we have verified through numerous simulations (not reported here due to space limitations) that any other choice of initial high (low) relay energy budgets naturally leads to similar transitory network behavior.

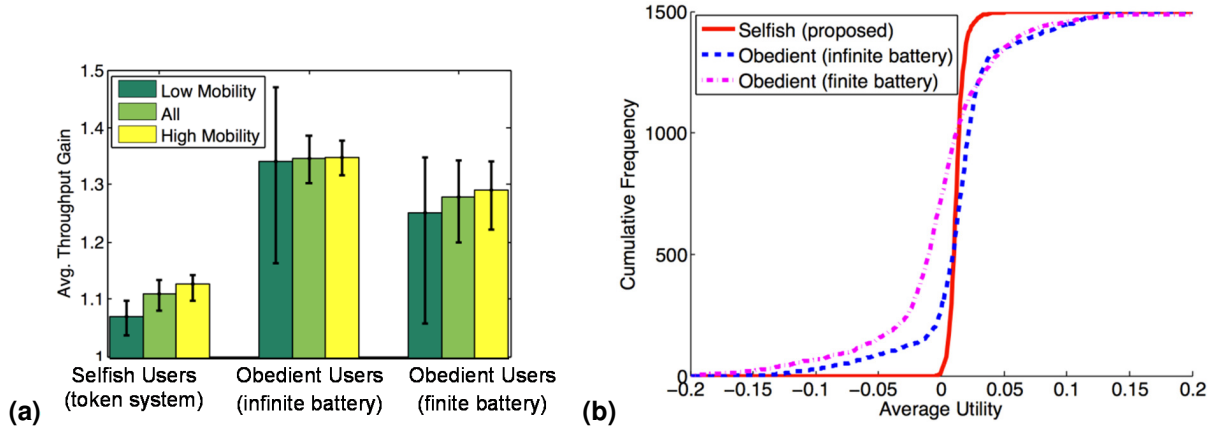

Fig. 5. Performance statistics for heterogeneous users (70% high mobility and 70% high relay energy budget) in three different scenarios. (Scenario 1): Self-interested users with finite relay energy budgets operating within the proposed token system. (Scenario 2): Obedient users with infinite relay energy budgets. (Scenario 3): Obedient users with finite relay energy budgets. (a) Average throughput gains. (b) Cumulative frequencies of users' average utilities.

In Fig. 5(a), we compare the average throughput gain obtained in the three aforementioned scenarios. In Fig. 5(b), we compare the cumulative frequencies of the users' average utilities for each of the scenarios in Fig. 5(a). From Fig. 5(a), it is clear that the *overall* network performance is best in the case when all users are obedient and have infinite relay energy budgets. However, Fig. 5(b) shows that over 250 of the *individual* UEs in the network (over 16%) actually obtain *negative* average utilities, and are therefore *being obliged to cooperate* even though it is not in their self-interest. These poor performing UEs typically have low mobility and are in the core of the network so they end up providing relay service much more frequently than they can exploit relay service themselves. These low mobility users are responsible for the relatively low $25^{th}$ percentile throughput gain illustrated in the center of Fig. 5(a). In contrast, the low mobility users at the periphery of the network tend to perform very well because they end up exploiting relay service more frequently than they need to provide it. In effect, the low mobility users at the periphery are *free-riders*: they benefit from cooperation without bearing its costs. These low mobility users are responsible for the relatively high $75^{th}$ percentile throughput gain illustrated in the center of Fig. 5(a) and are therefore largely responsible for the average network-wide throughput gains.

When all users are obedient, but have finite relay energy budgets, the *overall* network performance degrades relative to the infinite budget case [see Fig. 5(a)], but is still better than the case with self-interested users. The performance degradation relative to the infinite relay budget case occurs for two main reasons: (i) many low relay budget and low mobility users in the core of a cell provide relay service very frequently; consequently, they enter the dead state before making it to the periphery of a cell where they could balance out their average utility by exploiting relay service more frequently (as some low mobility users do in the infinite relay budget case); and (ii) as the low relay budget UEs enter the dead state, the density of active UEs in the network decreases leading to higher relay costs for all remaining active UEs. While the *overall* network performance is better than the case with self-interested users, Fig. 5(b) shows that over 700 of the *individual* UEs in the network (over 46%) actually obtain *negative* average utilities, and are therefore *being obliged to cooperate* even though it is not in their self-interest.

When all users are self-interested and have limited relay energy budgets, the overall network performance is lower than when all users are obedient [see Fig. 5(a)]; however, Fig. 5(b) shows that over 1490 of the individual UEs in the network (over 99%) obtain *positive* average utilities. This is because self-interested users will only relay if the marginal value of having an additional token outweighs the cost of obtaining it. Note that the few self-interested UEs that obtain (very small) negative average utilities each provided relay service a few more times than they exploited relay service just before the simulation ended; importantly, unlike in the cases with obedient users, the negative average utilities do not reflect a fundamental absence of incentive to relay, but are simply an artifact of the finite simulation duration.

*D. Impact of Mobility*

In this section, we use the simulation test-bed described in Section V.A to investigate how each UE's mobility impacts its own performance, the performance of other UEs, and the overall network performance. We consider five mobility mixtures in which the network comprises 10%, 30%, 50%, 70%, and 90% high mobility users, and the remaining users have low mobility. In total, there are $N = 1500$ users. We assume that all users start with the maximum relay energy budget, which is defined such that each user can R-ACK an average of 1000 relay requests before entering the dead state. We use a relatively high relay energy budget compared to the 3000 time slot simulation duration so we do not conflate the impact of low relay energy budgets with the impact of mobility (we investigate the impact of heterogeneous relay energy budgets in Section V.E). We use a discount factor $\beta = 0.995$.

In Fig. 6, we plot the average ORDR $\lambda$, average IRDR $\mu$, average RRE $e$, average relay lifetime (i.e., the amount of time a UE is not in the dead state), average R-ACK rate, and average throughput gain over 3000 time slots for users in each mobility class (i.e., high and low mobility) and for all users combined. The lower and upper error bars in Fig. 6 indicate the $25^{th}$ and $75^{th}$ percentiles, respectively, over the considered set of users.

**ORDR and IRDR:** Fig. 6(a) and Fig. 6(b) illustrate the average ORDR and average IRDR, respectively, under the various user mobility mixtures. Recall that, at the beginning of the simulation, users are uniformly distributed throughout the network. Since the low mobility users do not significantly deviate from their starting positions, and the high mobility users move throughout the network, the average ORDR and IRDR for users in each class is approximately the same; however, the variation of ORDRs and IRDRs across users within each class varies significantly. In particular, the low mobility users have a much larger range of ORDRs and IRDRs because, if they are at the periphery of a cell, then they will have large ORDRs and small IRDRs, and if they are in the core of a cell, then they will have small ORDRs and large IRDRs. In contrast, the high mobility users move between the peripheries and cores of various cells over time, and therefore experience less deviation in these parameters across the population.

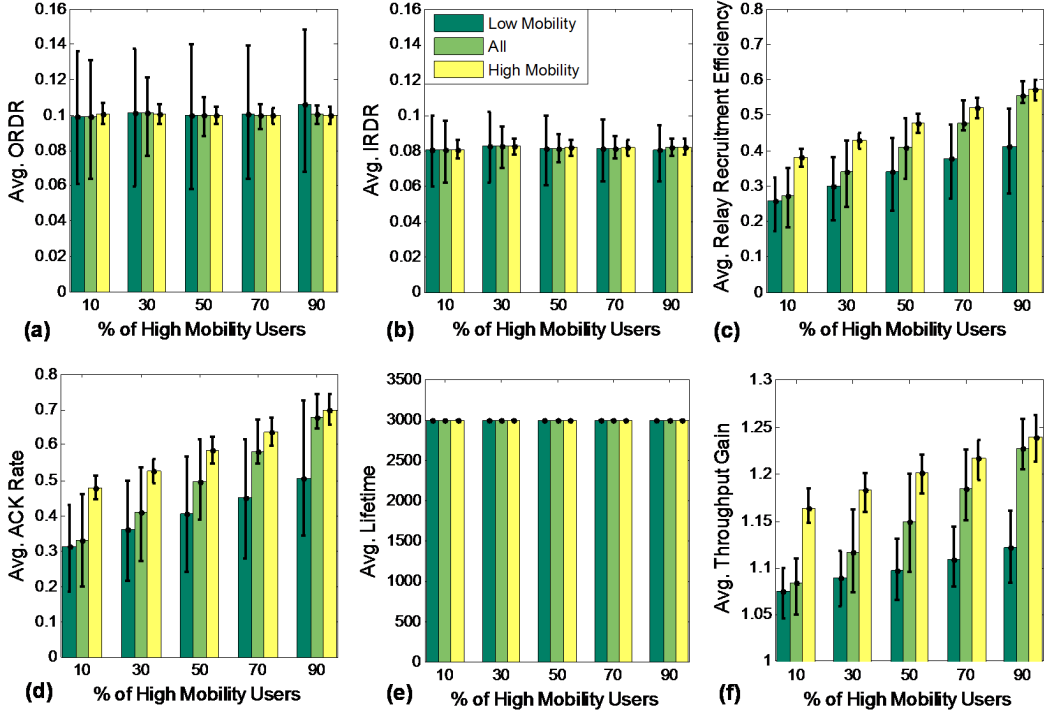

Fig. 6. Impact of the mobility mixture. (a) Average ORDR $\lambda$. (b) Average IRDR $\mu$. (c) Average RRE $e$. (d) Average R-ACK rate. (e) Average relay lifetime. (f) Average throughput gain.

**RRE and R-ACK rate:** Fig. 6(c) and Fig. 6(d) illustrate the average RRE and average R-ACK rate, respectively, under the various user mobility mixtures. Importantly, these parameters are intimately tied to each user's ORDR and IRDR. On average, low mobility users have lower relay recruitment efficiencies, lower R-ACK rates, and more variation of these parameters across the population. These effects emerge because low mobility users have strongly imbalanced ORDRs and IRDRs unless they are lucky enough to be situated between the core and periphery of a cell. If a UE's ORDR is larger than its IRDR, i.e., $\lambda > \mu$, then it tends to run out of tokens [i.e., $\mathbf{E}_k[I_{\{k>0\}}]$ in (19) will be small]. Without tokens, it cannot recruit relays, which reduces its RRE. Note that this also hurts other UEs because it reduces opportunities for them to earn tokens. On the other hand, if a UE's ORDR is smaller than its IRDR, i.e., $\lambda < \mu$, then it will tend to collect a surplus of tokens and not have any incentive to R-ACK incoming relay requests, which in turn reduces the RREs of the other UEs in the network.

**Average relay lifetime:** Fig. 6(e) illustrates the average relay lifetime of the UEs under various mobility mixtures. As we noted earlier, we intentionally selected the relay energy budgets such that no UEs would enter the dead state, i.e., $p = 0$, in these simulations.

**Throughput gain relative to direct transmission:** Fig. 6(f) illustrates the average throughput gain under the various user mobility mixtures. As expected, having a higher fraction of high mobility users in the network improves the average network throughput (relative to direct transmission only) because these users play different roles in the network over time, which enables a consistent exchange of tokens between UEs in the periphery and core of each cell.

*E. Impact of the Relay Energy Budget*

Users will select different relay energy budgets based on their preferences. In this section, we study how each UE's relay energy budget impacts its own performance, the performance of other UEs, and the overall network performance. We use the simulation test-bed described in Section V.A with discount factor $\beta = 0.995$. As in Section V.C, we assume that UEs with high (low) relay budgets can relay 100 (40) times on average before entering the dead state. We consider five mixtures in which the network comprises 10%, 30%, 50%, 70%, and 90% users with high relay energy budgets, and the remaining users have low relay energy budgets. In total, there are $N = 1500$ users. We assume that all users are high mobility users so we can clearly separate the impact of users with low relay energy budgets from the impact of users with low mobility. In Fig. 7, similar to Fig. 6, we plot the average ORDR $\lambda$, average IRDR $\mu$, average RRE $e$, average R-ACK rate, average relay lifetime, and average throughput gain for users in each relay energy budget class (i.e., high and low) and for all users combined. Note that these metrics are measured over the lifetime of each user (i.e., for all time slots before a user enters the dead state, $p = 0$). The total simulation duration is 3000 time slots. The error bars in Fig. 7 show the 25[th] and 75[th] percentiles over the considered set of users.

**ORDR and IRDR:** Fig. 7(a) and Fig. 7(b) illustrate the average ORDR and average IRDR, respectively, under the various relay energy budget mixtures. We observe that the relay energy budget has an insignificant impact on the ORDR and a minor impact on the IRDR of the high relay budget users. Specifically, as the number of UEs in the dead state increases, there are less UEs in the network that require help from relays, and therefore the high relay energy budget UEs' IRDRs decrease.

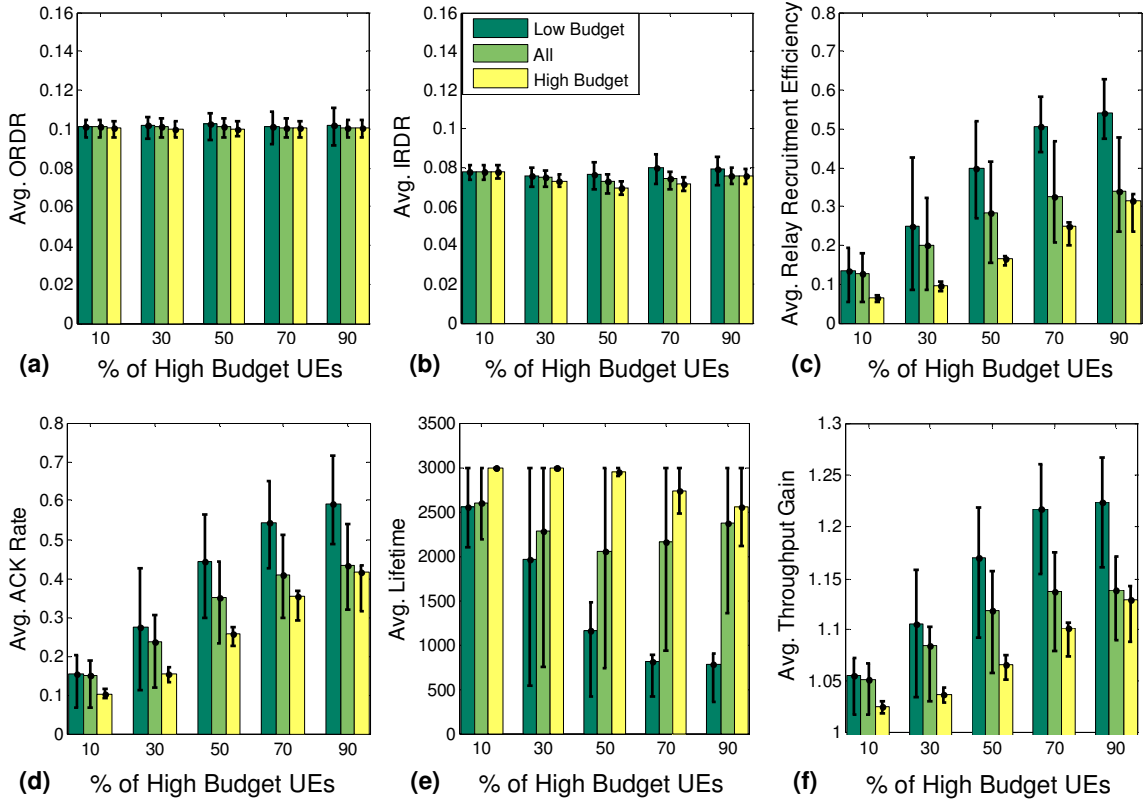

Fig. 7. Impact of different relay energy budget mixtures with $T = 9000$ tokens. (a) Average ORDR $\lambda$. (b) Average IRDR $\mu$. (c) Average RRE $e$. (d) Average R-ACK rate. (e) Average lifetime. (f) Average throughput gain.

**RRE and R-ACK rate:** Fig. 7(c) and Fig. 7(d) illustrate the average RRE and average R-ACK rate, respectively, under the various relay energy budget mixtures. When there are many low UEs with low relay budgets in the network, the average RRE and average R-ACK rate are low. This is because UEs with low relay budgets have lower thresholds in their optimal cooperation policies [see Fig. 4(a)]; therefore, when there are many low relay budget UEs, most of them will have sufficient tokens and, consequently, have no incentive to R-ACK incoming relay requests. Interestingly, when the low relay budget UEs carry *less* than the average number of tokens, the high relay budget UEs must carry *more* than the average number of tokens; therefore, most of the high relay budget UEs will also have sufficient tokens and, consequently, have no incentive to R-ACK incoming relay requests. Since the R-ACK rate is small, and the ORDR and IRDR are similar, it follows from the balance equation (19) that the RRE must also be small. Another interesting phenomenon in the network is that the UEs with high relay energy budgets have significantly lower average RREs and R-ACK rates than the UEs with low relay energy budgets. This is because, as the low relay budget UEs' drain their budgets, they further reduce their token holdings, which in turn increases the token holdings of UEs with high relay energy budgets. Since the high relay budget UEs are flooded with tokens, they have no incentive to R-ACK incoming relay requests. As illustrated in Fig. 8, these interesting phenomena can be compensated for by reducing the token supply in the network so that UEs hold fewer tokens and have incentive to R-ACK more incoming relay requests [28]. We further discuss the impact of the token supply on the overall network performance in Section V.F.

**Average relay lifetime:** Fig. 7(e) illustrates the average relay lifetime of the UEs under the various relay energy budget mixtures. The average relay lifetime (for UEs with both high and low relay budgets) decreases as the number of high relay budget UEs increases because the R-ACK rate is higher.

**Throughput gain relative to direct transmission:** Fig. 7(f) illustrates the average throughput gain under the various relay energy budget mixtures. The throughput gain is high (low) when the RRE and R-ACK rate are high (low).

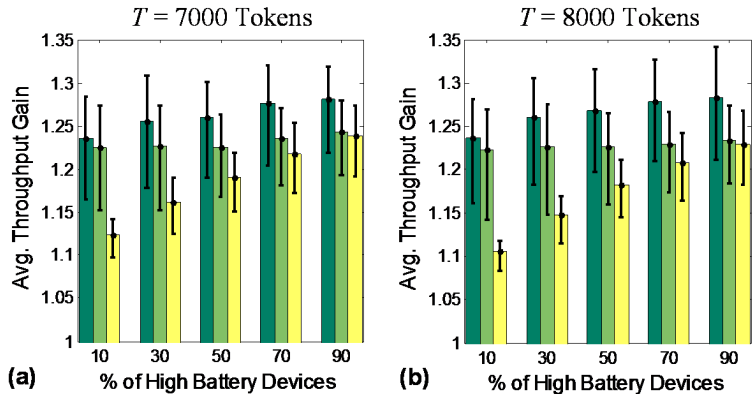

Fig. 8. Impact of token supply on the throughput gain under different relay energy budget mixtures. Compare to Fig. 7(f). (a) $T = 7000$ tokens. (b) $T = 8000$ tokens.

*F. Impact of the Token Supply*

The number of tokens in the network is an important design parameter that can significantly impact the average network throughput as illustrated in Fig. 8. In Fig. 9, we illustrate the impact of the token supply on the average throughput gain using token supplies ranging from 0 to 20000 in 1000 token increments. The results in Fig. 9 were generated under ideal conditions where all 1500 users have high mobility and high relay energy budgets; however, the general trend does not depend on the specific mixture of users. The best overall network performance in Fig. 9 is achieved when there are 8000 tokens in the network.

If there are too few tokens in the network, then many UEs will have zero tokens and therefore be unable to "buy" relay service when they need it. As illustrated in Fig. 9, this will significantly reduce the average network throughput. In the extreme case that there are zero tokens in the network (i.e., there is no token system), there is no incentive for the self-interested UEs to provide relay services, so the average network throughput will be equivalent to that of a network with only direct B2D transmission. If there are too many tokens in the network, then many UEs will have "sufficient" tokens (i.e., $k^j > K_{\text{th}}(p^j)$) and will therefore have no incentive to provide relay service (i.e., $\sigma^j(k^j, p^j) = 0$ ). Hence, as illustrated in Fig. 9, having too many tokens will also significantly reduce the overall network throughput.

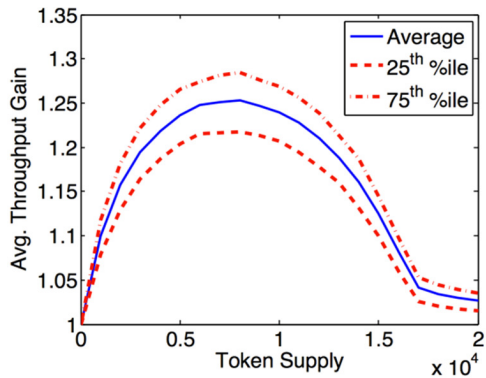

Fig. 9. Average throughput gain versus token supply.

## VI. CONCLUSION

We consider a cellular network in which token exchanges are used to provide self-interested users with incentive to relay data for other users. We formulate the decision problem faced by each UE (i.e., whether or not to acknowledge incoming relay requests) as a Markov decision process. We propose a supervised learning algorithm that a UE can deploy to learn its optimal cooperation policy online based on its experience. We evaluate the proposed learning algorithm in various deployment scenarios involving both high mobility and low mobility users and UEs with both high and low relay energy budgets. Our simulation results indicate that individual UEs have the greatest incentive to cooperate when (i) the network contains many UEs with high relay energy budgets, (ii) many highly mobile users (e.g., users in motor vehicles), and (iii) there are neither too few nor too many tokens in the network. Moreover, our simulation results demonstrate that, in contrast to the conventional approach to cooperation in which users are obedient and are obliged to cooperate, the proposed solution protects individual users from providing relay service significantly more than they receive relay service, thereby preventing users from ending up with negative average utilities.

In future work, we plan to investigate (i) how the token supply can be dynamically adapted to maximize the average network throughput; (ii) how the token system can be adapted to work with cooperation schemes that use multiple relays; and (iii) how heterogeneous networks with Wi-Fi offloading, macrocells, and small cells can improve the performance of the considered token system, and vice versa.